\documentclass{aa}
\usepackage{graphicx}
\usepackage{txfonts}

\newcommand{\msun}{\mbox{${\rm M}_\odot$}}

\newcommand{\kms}{\mbox{${\rm km~s}^{-1}$}}
\newcommand{\ergs}{\mbox{${\rm erg~s}^{-1}$}}
\newcommand{\ergcms}{\mbox{${\rm erg~cm}^{-2}~s^{-1}$}}

\begin{document}
%\headnote{Research Note}

\title{Catalogue of high-mass X-ray binaries in the Galaxy ($4^{th}$ edition)
\thanks {Table 1 and the references are only available in electronically form at the CDS via anonymous
ftp (130.79.128.5) or via http://cdsweb.u-strasbg.fr/cgi-bin/qcat?J/A+A/???/????} }
\author{Q.Z. Liu\inst{1, 2}, J. van Paradijs\inst{2}, and E.P.J. van den Heuvel\inst{2}}

\offprints{Q.Z. Liu}
%\offprints{Q.Z. Liu (email: qzliu@)}

\institute{Purple Mountain Observatory, Chinese Academy of Sciences, Nanjing 210008, P.R. China\\
           \email{qliu@astro.uva.nl}
 \and
     Sterrenkundig Instituut "Anton Pannekoek", Universiteit van Amsterdam, Kruislaan 403,
                    1098 SJ Amsterdam, The Netherlands}

\date{Received date/ Accepted date}

\abstract  % context heading (optional)
  {}
  {The aim of this catalogue is to provide the reader with some basic information on the X-ray sources and their
  counterparts in other wavelength ranges ($\gamma$-rays, UV, optical, IR, radio). }
{Some sources, however, are only tentatively identified as high-mass X-ray binaries on the basis of their X-ray
properties similar to the known high-mass X-ray binaries. Further identification in other wavelength bands is needed to
finally determine the nature of these sources. In cases where there is some doubt about the high-mass nature of the
X-ray binary this is mentioned. Literature published before 1 October 2005 has, as far as possible, been taken into
account.}
{We present a new edition of the catalogue of high-mass X-ray binaries in the Galaxy. The catalogue contains source
name(s), coordinates, finding chart, X-ray luminosity, system parameters, and stellar parameters of the components and
other characteristic properties of 114 high-mass X-ray binaries, together with a comprehensive selection of the
relevant literature. About 60\% of the high-mass X-ray binary candidates are known or suspected Be/X-ray binaries,
while 32\% are supergiant/X-ray binaries.}
 {}

%\abstract{We present a new edition of the catalogue of high-mass X-ray binaries in the Galaxy. The catalogue contains
%source name(s), coordinates, finding chart, X-ray luminosity, system parameters, and stellar parameters of the
%components and other characteristic properties of 114 high-mass X-ray binaries, together with a comprehensive selection
%of the relevant literature. The aim of this catalogue is to provide the reader with some basic information on the X-ray
%sources and their counterparts in other wavelength ranges ($\gamma$-rays, UV, optical, IR, radio). About 60\% of the
%high-mass X-ray binary candidates are known or suspected Be/X-ray binaries, while 32\% are supergiant/X-ray binaries.
%Some sources, however, are only tentatively identified as high-mass X-ray binaries on the basis of their X-ray
%properties similar to the known high-mass X-ray binaries. Further identification in other wavelength bands is needed to
%finally determine the nature of these sources. In cases where there is some doubt about the high-mass nature of the
%X-ray binary this is mentioned. Literature published before 1 October 2005 has, as far as possible, been taken into
%account.


\maketitle
\authorrunning{Q.Z. Liu et al.}
\titlerunning{Catalogue of high-mass X-ray binaries}

\section{Introduction}

High-mass X-ray binaries (HMXBs) were among the very first bright X-ray sources detected and optically identified in
the 1970s. The systems comprise a compact object orbiting a massive OB class star. The compact object should be either
a neutron star (NS) or black hole and is a strong X-ray emitter via the accretion of matter from the OB companion.
Thus, cataclysmic variables do not belong to HMXBs (for a catalogue of cataclysmic variables see Ritter \& Kolb 2003).
Conventionally, HMXBs can be further divided into two subgroups: those in which the primary is a Be star (Be/X-ray
binary) and those in which the primary is a supergiant (SG/X-ray binary). One can refer to the book by Lewin \& van der
Klis (2006) for a full understanding of the various aspects of HMXBs.

The majority of the known high-mass X-ray binaries are Be/X-ray systems (BeXRBs), especially those in the Magellanic
Clouds (Liu, van Paradijs \& van den Heuvel 2005). Meurs \& van den Heuvel (1989) predict 2000--20000 Be/X-ray binaries
in the Galaxy. In Be systems, the compact object is a neutron star and is typically in a wide, moderately eccentric
orbit, and it spends little time in close proximity to the dense circumstellar disk surrounding the Be companion (Coe
2000; Negueruela 2004). No black hole and Be star system has been found yet (see Zhang, Li \& Wang 2004). X-ray
outbursts will be expected when the compact object passes through the Be-star disk, accreting from the low-velocity and
high-density wind around Be stars, and thus collectively termed Be/X-ray transients. Their X-ray spectra are usually
hard. The hard X-ray spectrum, along with the transience, is an important characteristic of the Be/X-ray binaries.

In the second group of HMXB systems, the compact star orbits a supergiant early-type star, deep inside the highly
supersonic wind (for a review see Kaper et al. 2004). The X-ray luminosity is either powered by the strong stellar wind
of the optical companion or Roche-lobe overflow. In a wind-fed system, accretion from the stellar wind results in a
persistent X-ray luminosity of $10^{35}$--$10^{36}$ \ergs, while in a Roche-lobe overflow system, matter flows via the
inner Lagrangian point to an accretion disc. A much higher X-ray luminosity ($\sim10^{38}$ \ergs) is produced.

Five years after the publication of the previous (3rd) edition (Liu, van Paradijs \& van den Heuvel 2000), the amount
of new literature and the number of new objects to be included have again grown so much that it seems worthwhile to
concentrate the information of HMXBs in the Magellanic Clouds and in the Galaxy separately. In a previous paper we
published a catalogue of 128 HMXBs in the Magellanic Clouds (Liu, van Paradijs \& van den Heuvel 2005), and here we
present a catalogue of HMXBs in the Galaxy. We briefly recall some of the developments that, over the past five years,
have had (and still have) a major impact on this catalogue.

Due to the much increased sensitivity and spatial resolution achievable with the Chandra and the XMM-Newton X-ray
observatories, as well as with the Hubble Space Telescope and large ground-based radio telescopes, more accurate
positions of X-ray binaries have been determined, resulting in the unambiguous discovery of the optical and/or IR
counterpart to some X-ray sources. Moreover, the number of HMXBs in external galaxies is also rapidly increasing, e.g.,
the X-ray binaries in M81 (Swartz et al. 2003) and the Antennae galaxies (Zezas et al. 2002). Most of the
ultra-luminous X-ray sources in starburst and spiral galaxies (Liu \& Mirabel 2005) are believed to be HMXBs with a
black hole, specifically the B0 Ib star to NGC 5204 X-1 (Liu et al. 2004), the O9I star to M 33 X-7 (Pietsch et al.
2004) and the mid-B supergiant to M101 ULX-1 (Kuntz et al. 2005). It has been proposed that the collective X-ray
luminosity of high-mass X-ray binaries can be used as an indicator of the star-formation rate for the host galaxy
(Grimm, Gilfanov \& Sunyaev 2003).

Since its launch in 2002, INTEGRAL has been revealing hard X-ray sources that were not easily detected in earlier soft
X-ray (typically $\leq$10 keV) observations. Hard X-rays are not easily absorbed by matter and thus are highly
penetrating. Such radiation is, therefore, ideal for probing high-energy emitting sources in dense regions. INTEGRAL
has been performing a regular survey of the Galactic plane and a deep exposure of the Galactic Center as part of its
Core Program (Winkler et al. 2003). A group of hard X-ray sources emitting in the hard X-ray and soft $\gamma$-ray
regions have been discovered in the course of the INTEGRAL observations, which are highly absorbed, i.e., with column
densities higher than about 10$^{23}$\,cm$^{-2}$. The X-ray, as well as the optical/IR, properties of these sources and
their location in the sky suggest that they may belong to the class of high-mass X-ray binaries, some of them possibly
long-period X-ray pulsars. The donors in these binaries are most probably giant or supergiant stars (see Kuulkers
2005). For the details of all INTEGRAL sources please refer to the web page of Jerome Rodriguez
(http://isdc.unige.ch/$^\sim$rodrigue/html/igrsources.html).

In recent years there has been growing evidence that there is a class of X-ray binaries referred to as fast X-ray
transients, characterized with X-ray outburst durations on the order of a few hours and peak fluxes on the order of
$10^{-9}\,$\ergcms (2--10 keV). They are not readily identified with other similarly active X-ray sources: magnetically
active nearby stars (i.e., DY Dra, RS CVn, or pre-main sequence stars) or superbursters (e.g., Kuulkers 2004). These
sources lie in the vicinity of the Galactic Center, a region extensively monitored by INTEGRAL and other satellites.
The high fluxes and the lack of nearby counterparts suggest high luminosities that would indicate an X-ray binary
origin. Several of these fast transients are associated with OB supergiants (Coe et al. 1996; Halpern et al. 2004;
Negueruela et al. 2006; Smith et al. 2006; Pellizza et al. 2006), so they are suggested to be a new class of high-mass
X-ray binaries (Negueruela et al. 2005).

%As a result of systematic searches for potential double degenerate Type Ia supernova
%progenitors (see e.g. ) the number of detached short-period double white dwarf systems
%has increased dramatically and is likely to continue to grow substantially in the future.

The aim of our catalogue is to provide some basic information on the X-ray sources and their counterparts, as well as
the binary properties of the system in question, and easy access to the recent literature. No attempt has been made to
compile complete reference lists. Much effort has been made to avoid errors and to keep the information up to date.
Nevertheless, the authors are well aware that this edition too may contain errors and may be
incomplete. %It is certainly incomplete with respect to the references quoted.

%All the tabular material contained in this catalogue is published in electronic form only. It
%is available in electronic form at the CDS via anonymous ftp to cdsarc.u-strasbg.fr
%(130.79.128.5) or via http://cdsweb.u-strasbg.fr/cgi-bin/qcat?J/A+A/404/301.

%In addition to the electronic version we provide postscript files for a printable stand alone
%version at astro-ph.

\section{Description of the table}

Table 1 lists the 114 HMXB candidates in the Galaxy. The format of Table 1 is similar to the previous edition and
almost the same with the HMXB catalogue in the Magellanic Clouds (Liu, van Paradijs \& van den Heuvel 2000; 2005), of
which the present catalogue is meant to be an update. If there is any doubt about an entry, a question mark follows the
item. In the table the sources are ordered according to the right ascension of sources; part of the (mainly numerical)
information on a source is arranged in six columns, below which additional information is provided for each source in
the form of key words with reference numbers [in square brackets]. The columns have been arranged as follows.

In Column 1 the first line contains the source name, with rough information on its sky location, according to the
conventional source nomenclature of space experiments in which the source was detected, hhmm$\pm$ddd or
hhmm.m$\pm$ddmm. Here hh, mm, and ss indicate the hours, minutes and seconds of right ascension, ddd the declination in
units of 0.1 degree (in a small number of cases, the coordinates shown in the name are given with more, or fewer,
digits). The prefix J indicates a name based on J2000 coordinates. Otherwise, 1950 coordinates were used in the name.
An alternative source name is given in the second line. In the third line of Column
1, the source types are indicated with a letter code, as follows:\\
\\
$\bullet$ P: X-ray pulsar (66);\\
$\bullet$ T: transient X-ray source (62);\\
$\bullet$ E: eclipsing system (9);\\
$\bullet$ R: radio emitting HMXBs (9);\\
$\bullet$ C: cyclotron resonance scattering feature at X-ray

~~spectrum (16);\\
$\bullet$ U: ultra-soft X-ray spectrum (1). These sources include

~~black-hole candidates (BHC); some $`$extreme ultra-

~~soft$'$ (EUS) source may be a white dwarf (WD) on

~~whose surface steady nuclear burning takes place.\\

Column 2 contains the right ascension (RA) and declination (DEC) of the source for equinox J2000.0 in the first two
lines. RA is given as hhmmss.s to an accuracy of 0.1 s, DEC is given in $^\circ$ ' ", to an accuracy of 1" (in a small
number of cases, the coordinates are given with more, or fewer, digits). The third line gives the galactic longitude
and latitude to an accuracy of 0.1$^\circ$ (except for sources close to the galactic center, where these coordinates
are given to 0.01$^\circ$). A reference for the source position is given below the columnar information under $'pos.'$.
In the parentheses following the $'pos.'$, we provide some information on the type of observation from which the source
position has been derived. The following abbreviations are used: o, optical; x, X-ray; r, radio; IR, infrared.
Following the type of observation, we give an indication of the accuracy of this position, in the form of equivalent
(90 percent confidence level) error radii, but in several cases this can only be considered an approximation (e.g. when
the error box is not circular). When no accuracy is quoted, it is about one arcsecond or better.

The first and second lines of Column 3 give the names of the optical counterpart to an X-ray source. The third line
contains a reference to a finding chart. An asterisk followed by a number or letter refers to the star number used in
the finding chart; "star" refers to a star in the finding chart that has not been assigned a number or letter. Many
optical counterparts have been indicated with a variable-star name, as given in the
$General~Catalogue~of~Variable~Stars$ and in recent name lists of variable stars as published regularly in the
$IAU~Information~Bulletin~on~Variable~Stars$, or a number in a
well-known catalogue (e.g., HD, SAO, GSC, 2MASS). %For X-ray sources in globular clusters, the
%cluster name is here given, in addition to the name of a stellar optical counterpart.

The fourth column contains some photometric information on the optical counterpart. In the first line, the apparent
visual magnitude, $V$, and the color indices $B-V$ and $U-B$ are listed. The second line contains the spectral type of
the optical counterpart and an estimate of the interstellar reddening, $E_{B-V}$.

The fifth column lists the near-infrared magnitudes at J (1.25 microns), H (1.65 microns), and Ks (2.17 microns),
detected with the Two Micron All Sky Survey (2MASS)(Skrutskie et al. 2006).

In Column 6, the maximum X-ray flux, or the range of observed X-ray fluxes (2--10 keV, unless otherwise indicated), is
given in units of
\begin{eqnarray*}
1\mu Jy & = &10^{-29}~ erg~ cm^{-2}~ s^{-1} Hz^{-1}\\
       & = & 2.4\times 10^{-12}~ erg~  cm^{-2}~  s^{-1}~ keV^{-1}.
\end{eqnarray*}

%1 mCrab= 1.4x10^{-11} erg/s/cm2 for a source with a Crab-like spectrum.

The first line in Column 7 gives the orbital period in days. The second line contains the pulse period for X-ray
pulsars, in seconds. The third line contains a reference in which the orbital and/or pulse periods were detected.

\section{Conclusions and remarks}
The current version of this catalogue provides tabulated data and references for 114 objects, including 35 newly
discovered HMXBs (2 previously listed in our low-mass X-ray binary catalogue), as well as 79 $``$old" ones listed in
the previous catalogue. Compared with the 3rd edition, the number of HMXBs in the Galaxy listed has increased by
$\sim$43\%. Among the 114 HMXB candidates, we find 39(13) confirmed(suspected) Be/X-ray binaries (another 4 dubious
Be/X-ray binaries, probably a white dwarf with a Be companion), 18(11) SG/X-ray binaries, and 6 other sources with
peculiar features (CI Cam, IGR J16318-4848, Cyg X-3, LS 5039, V4641 Sgr, and SS 433). The remaining 23 systems are
candidates of HMXBs. No detailed information on optical companions is known.

We would like to make some remarks on several individual sources. Both KS 1947+300 and GRO J1948+32 have been listed in
the 3rd edition. Swank \& Morgan (2000), however, found that the transient X-ray source KS 1947+300 has pulsations that
are identical to the pulsar GRO J1948+32, and concluded that they are the same source. Also, the ROSAT source, RX
J1037.5$-$5647, is the only one in the Uhuru error boxes 4U 1036$-$56 (Motch et al. 1997), so probably having the same
origin.

MXB 0656$-$072 was described as a soft X-ray transient during the 1975 outburst (Clark 1975) and previously catalogued
as a low-mass X-ray binary. However, observations with the PCA on board the RXTE satellite in October 2003 indicated
that the source is actually a hard X-ray pulsar with a spin period of 160.7 s (Morgan, Remillard \& Swank 2003).
Moreover, the position of the ROSAT counterpart, RX J065817.7$-$071228, is consistent with a variable-reddened O9.7Ve
spectral type Be star, suggesting a Be X-ray binary nature (Pakull, Motch \& Negueruela 2003). Another source, SAX
J1819.3$-$2525/V4641 Sgr, was previously classified as a low-mass X-ray binary. Although the accretion processes in
this system are typical of those seen in LMXBs, recent spectroscopic observations yielded a mass of the compact star in
the range 5.49--8.14 \msun, and the optical companion was found to be a B9III star (Orosz et al. 2001). Chaty et al.
(2003) also showed that the stellar type of the companion star was early type between B3 and A2 and that the binary was
between intermediate-mass X-ray binary and HMXB. Therefore, this system is more consistent with a HMXB than a LMXB.

We also wish to emphasize that some sources listed in this catalogue are still uncertain. They should be considered
with caution, in view of all the further work needed. Some sources are tentatively classified as massive X-ray binaries
due to their X-ray properties, for instance, a hard X-ray transient, a fast X-ray transient, or an extremely absorbed
X-ray source. No counterpart at other bands has been found or no pulse period has been detected, or both. Also, the
compact object in some weak or soft X-ray sources may be a white dwarf instead of a neutron star, e.g., 1H 0556+286, 1H
0749-600, 1H 1249-637, and 1H 1253-761 (Torrejon \& Orr 2001 and see Table 1). If this is the case, they should be
excluded from this catalogue.

IGR J12349$-$6434 is tentatively identified as a HMXB candidate, primarily due to an early-type star, RT Cru, in the
error of the X-ray source (Masetti et al. 2005), while XTE J1859+083 is included simply due to the detection of X-ray
pulsations (Marshall et al. 1999). Highly absorbed hard X-ray sources are suggested as being high-mass X-ray binaries
(see Kuulkers 2005). Two hard X-ray sources, IGR J16418$-$4532 and IGR J16493$-$4348, with column densities of about
10$^{23}$\,cm$^{-2}$ (Walter et al. 2004; Markwardt et al. 2005), are also included in this catalogue. Fast X-ray
transients are probably another new class of high-mass X-ray binaries with a supergiant companion (Smith et al. 2006;
Negueruela et al. 2006). The two fast transients, AX J1749.1$-$2733 (Sakano et al. 2002, see in't Zand 2005) and XTE
J1901+014 (Remillard \& Smith 2002), together with 8 other members (XTE J1739-302, IGR J17544-2619, IGR J16465-4507, AX
1845.0-0433, AX J1841.0-0536, SAX J1818.6-1703, IGR J16479-4514, and IGR J11215-5952), are included in this edition.

Finally, the following four sources are no longer listed in our catalogue. Now it is quite clear that they do not
belong to HMXBs.

1E 1024.0$-$5732/Wack 2134 was discovered with the Einstein Observatory. Caraveo et al. (1989) suggested that a highly
reddened O5 star was the most likely optical counterpart, and therefore proposed the system was a HMXB. The source was
listed in the previous catalogue. Optical and X-ray data, however, favor the system containing a Wolf-Rayet star and an
O-type star (Mereghetti et al. 1994; Reig 1999). The origin of the X-rays from this source is explained by the
colliding-wind binary model (Reig 1999).

1E 1048.1$-$5937 and GS 1845$-$03 were listed in the previous catalogue. GS 1845$-$03 is, however, probably the same as
the ASCA source, AX J1845.0$-$0258. Both the compact star of 1E 1048.1$-$5937 and AX J1845.0$-$0258 (Torii et al. 1998;
Gotthelf \& Vasisht 1998) very likely belong to the "anomalous" X-ray pulsars, which are now thought to be related to
the magnetars. Mereghetti et al. (1998) argued that if 1E 1048.1$-$5937 has a companion this star must be either a
white dwarf or a helium-burning star instead of a main-sequence companion.

The peculiar galactic X-ray source, 4U 1954+319, has been listed in the previous editions as an HMXB. Masetti et al.
(2006), however, found that the suspected field M-type giant star is indeed the counterpart of the X-ray source, based
on the Chandra Observation. They suggest that 4U 1954+319 is a wide-orbit LMXB, most likely a neutron star, accreting
from the wind of an M-type giant.

V669 Cep used to be identified as the optical counterpart to the hard X-ray source, 1H 2214+589. However, in't Zand et
al. (2000) identify another Be star as the optical counterpart to its BeppoSAX counterpart, SAX J2239.3+6116. Instead,
V669 Cep is identified as the optical counterpart to RX J2226.6+6113 (Halpern et al. 2001). Both Hang, Liu \& Xia
(1999) and Halpern et al. (2001) classify V669 Cep as a Herbig Ae/Be star, while Miroshnichenko et al. (2002) suggest
it as a binary, consisting of a hot, low luminosity B4--B6 star and a cool companion, most likely a late-type giant. We
have excluded this source in the 4th edition, since in both cases RX J2226.6+6113/V669 Cep is unlikely to be an HMXB,
although it is still a HMXB in SIMBAD.

\textbf{Noted added in proof}: After we submitted the paper, we noticed that Swift J1626.6-5156 (Negueruela et al.
2006, ATel 739), XTE J1716-389 (Cornelisse et al. 2006, MNRAS, 366, 918), IGR J16207-5129, and AX J1700.2-4220
(Palazzi, et al. 2006, ATel 783) are likely candidates for high-mass X-ray binaries.

\acknowledgements We wish to thank Marc Rib\'o, Sylvain Chaty  %, Deepto Chakrabarty, Nicola Masetti,
 and Ignacio Negueruela for carefully reading the manuscript and for their useful comments. We also thank
Leonardo Pellizza and Sylvain Chaty %, Jerome Rodriguez,
 for providing us with information on
IGR J17544$-$2619 prior to publication. Finally, we are very grateful to the referees, Malcolm Coe and Sebastien
Derriere, for their careful and thorough reading of this paper.

This research has made use of the SIMBAD data base operated at the CDS, Strasbourg, France, and NASA's Astrophysics
Data System (ADS). This publication also makes use of data products from the Two Micron All Sky Survey, which is a
joint project of the University of Massachusetts and the Infrared Processing and Analysis Center/California Institute
of Technology, funded by the National Aeronautics and Space Administration and the National Science Foundation.

 QZL is partially supported by the Royal
Science Foundation of The Netherlands, the National Natural Science Foundation of China under Grants 10173026 and
10433030, and the Major State Basic Research Development Program of China (973 Program) under Grant G1999075405.

%We wish to thank H.-C. Thomas and V. Burwitz for keeping us informed about the latest results
%regarding the optical identification and follow-up observations of new CVs from the ROSAT All
%Sky Survey. We also thank J. Thorstensen for providing us with information on numerous new CVs
%with newly measured periods prior to publication.

\clearpage
\begin{table*}[t]
\caption{High-mass X-ray binaries in the Galaxy}
\begin{tabular}{p{3.6cm}p{1.9cm}p{2.3cm}p{3.3cm}p{0.8cm}p{1.9cm}p{1.7cm}}
%\noalign{\smallskip}
\hline \noalign{\smallskip}
%\multicolumn{7}{p{17.6cm}}{Table 3.  (continued) }\\
%\hline \noalign{\smallskip}
Name(s)  & RA (J2000)       & Opt. Ctp. & V, $B-V$, $U-B$     & J & F$_{x}$ & P$_{orb}$(d)   \\
         & DEC (J2000)      &           & Sp. type, E$_{B-V}$ & H & $\mu$Jy & P$_{pulse}$(s) \\
type     & $l^{II}, b^{II}$ & [FC]      &                     & K &         &                \\

\noalign{\smallskip}\hline \noalign{\smallskip}
IGR J00370+6122       & 00 37 10      & BD +60 73     & 9.65, 0.56,       & 8.389 & 8           & 15.665 \\
1RXS J003709.6+612131 & +61 21 35     & LS I +61 161  & B0.5II--III, 0.75 & 8.265 & (2--28 keV) &       \\
                      & 121.2, $-$1.5 &               & [640]             & 8.166 & [165]       & [165] \\
\\
\multicolumn{7}{p{17.5cm}}{$Pos$. (o): [640]; distance of 3.0 kpc: [640]; variable hard X-ray source: [165]; no radio
emission detected: [75].
 }\\
\hline
\end{tabular}
\end{table*}

\clearpage
\begin{table*}[t]
\begin{tabular}{p{3.2cm}p{1.9cm}p{2.3cm}p{3.5cm}p{0.8cm}p{1.8cm}p{2.0cm}}
\noalign{\smallskip}
\multicolumn{7}{p{8.6cm}}{Table 1.  (continued) }\\
%\noalign{\smallskip}
\hline \noalign{\smallskip}
Name(s)    & RA (J2000)        & Opt. Ctp. & V, $B-V$, $U-B$     & J & F$_{x}$  & P$_{orb}$(d)   \\
           & DEC (J2000)       &           & Sp. type, E$_{B-V}$ & H & $\mu$Jy  & P$_{pulse}$(s) \\
 Type      & $l^{II}, b^{II}$  & [FC]      &                     & K &          &                \\

\noalign{\smallskip}\hline \noalign{\smallskip}
1E 0236.6+6100 & 02 40 31.7  & LS I +61$^\circ$303 & 10.7, 0.8, $-$0.3 & 8.621 & 0.2          & 26.496      \\
               & +61 13 46   & V615 Cas            & B0Ve, 0.75        & 8.228 & (0.2--5 keV) &             \\
R              & 135.7, +1.1 & [278]               & [179, 316, 726]   & 7.917 & [43]         & [755, 254] \\
\\
\hline
\end{tabular}
\end{table*}

\clearpage
\begin{table*}[t]
\begin{tabular}{p{3.3cm}p{1.9cm}p{2.3cm}p{3.3cm}p{0.8cm}p{1.9cm}p{2.0cm}}
\noalign{\smallskip}
\multicolumn{7}{p{8.6cm}}{Table 1.  (continued) }\\
\hline \noalign{\smallskip}
Name(s)   & RA (J2000)       & Opt. Ctp. & V, $B-V$, $U-B$     & J & F$_{x}$  & P$_{orb}$(d)   \\
          & DEC (J2000)      &           & Sp. type, E$_{B-V}$ & H & $\mu$Jy  & P$_{pulse}$(s) \\
 Type     & $l^{II}, b^{II}$ & [FC]      &                     & K &          &                \\

\noalign{\smallskip} \hline \noalign{\smallskip}
1H 0556+286   & 05 55 55.1     & HD 249179  & 9.2,~~ ,   & 9.366 & 1.1      &        \\
              & +28 47 06      &            & B5ne,      & 9.238 &          &        \\
              & 181.3, +1.9    &            & [608, 778] & 9.068 & [850]    &        \\
\\
\hline
\end{tabular}
\end{table*}

\clearpage
\begin{table*}[t]
\begin{tabular}{p{3.3cm}p{1.9cm}p{2.3cm}p{3.3cm}p{0.8cm}p{1.9cm}p{2.0cm}}
\noalign{\smallskip}
\multicolumn{7}{p{8.6cm}}{Table 1.  (continued) }\\
\hline \noalign{\smallskip}
Name(s) & RA (J2000)        & Opt. Ctp. & V, $B-V$, $U-B$     & J & F$_{x}$ & P$_{orb}$(d)   \\
        & DEC (J2000)       &           & Sp. type, E$_{B-V}$ & H & $\mu$Jy & P$_{pulse}$(s) \\
Type    & $l^{II}, b^{II}$  & [FC]      &                     & K &         &                \\

\noalign{\smallskip} \hline \noalign{\smallskip}
GS 0834$-$430 & 08 35 55.4     &  $^*$D  & 20.4, ~~,          & & 30--300 & 105.8        \\
              & 43 11 11.9     &         & B0--2 III--Ve, 4.0 & &         & 12.3         \\
T, P          & 177.7, 36.7    & [349]   & [349]              & & [249]   & [249, 836] \\
\\
\hline

\end{tabular}
\end{table*}

\clearpage
\begin{table*}[t]
\begin{tabular}{p{3.3cm}p{1.9cm}p{2.3cm}p{3.3cm}p{0.8cm}p{1.9cm}p{2.0cm}}
\noalign{\smallskip}
\multicolumn{7}{p{8.6cm}}{Table 1.  (continued) }\\
\hline \noalign{\smallskip}
Name(s)   & RA (J2000)       & Opt. Ctp. & V, $B-V$, $U-B$     & J & F$_{x}$ & P$_{orb}$(d)   \\
          & DEC (J2000)      &           & Sp. type, E$_{B-V}$ & H & $\mu$Jy & P$_{pulse}$(s) \\
 Type     & $l^{II}, b^{II}$ & [FC]      &                     & K &         &                \\

\noalign{\smallskip} \hline \noalign{\smallskip}
2S 1145$-$619  & 11 48 00.0     & V801 Cen     & 9.3, 0.18, $-$0.81 & 8.677 & 4--1000  & 187.5      \\
2E 1145.5-6155 & $-$62 12 25    & HD 102567    & B0.2IIIe, 0.35     & 8.577 &          & 292.4      \\
P, T           & 295.6, $-$0.2  & [65]         & [568, 564, 816]    & 8.403 & [64]     & [547, 819] \\
\\
\hline
\end{tabular}
\end{table*}

\clearpage
\begin{table*}[t]
\begin{tabular}{p{3.3cm}p{1.9cm}p{2.3cm}p{3.3cm}p{0.8cm}p{1.9cm}p{2.0cm}}
\noalign{\smallskip}
\multicolumn{7}{p{8.6cm}}{Table 1.  (continued) }\\
\hline \noalign{\smallskip}
Name(s)   & RA (J2000)       & Opt. Ctp.  & V, $B-V$, $U-B$     & J & F$_{x}$  & P$_{orb}$(d)   \\
          & DEC (J2000)      &            & Sp. type, E$_{B-V}$ & H & $\mu$Jy  & P$_{pulse}$(s) \\
type      & $l^{II}, b^{II}$ & [FC]       &                     & K &          &                \\

\noalign{\smallskip} \hline \noalign{\smallskip}
1H 1255$-$567 & 12 54 36.9   & $\mu^2$ Cru    & 5.17, $-$0.12, $-$0.51 & 5.316 & 0.8      &        \\
              & $-$57 10 07  & SAO 240367     & B5 Ve, 0.04            & 5.383 &          &        \\
              & 303.4, +5.7  &                & [564, 710]             & 5.313 & [850]    &        \\
\\
\hline
\end{tabular}
\end{table*}

\clearpage
\begin{table*}[t]
\begin{tabular}{p{3.3cm}p{1.9cm}p{2.3cm}p{3.3cm}p{0.8cm}p{1.9cm}p{2.0cm}}
\noalign{\smallskip}
\multicolumn{7}{p{8.6cm}}{Table 1.  (continued) }\\
\hline \noalign{\smallskip}
Name(s)    & RA (J2000)       & Opt. Ctp. & V, $B-V$, $U-B$     & J & F$_{x}$  & P$_{orb}$(d)   \\
           & DEC (J2000)      &           & Sp. type, E$_{B-V}$ & H & $\mu$Jy  & P$_{pulse}$(s) \\
 Type      & $l^{II}, b^{II}$ & [FC]      &                     & K &          &                \\

\noalign{\smallskip} \hline \noalign{\smallskip}
1H 1555$-$552 & 15 54 21.8    & HD 141926      & 8.60, 0.56, $-$0.43 & 7.098 & 1.7       &      \\
              & $-$55 19 45   & SAO 243098     & B2IIIn,             & 6.841 &           &      \\
              & 327.0, $-$1.2 & [719]          & [192, 806]          & 6.509 & [850]     &      \\
\\
\hline
\end{tabular}
\end{table*}

\clearpage
\begin{table*}[t]
\begin{tabular}{p{3.3cm}p{1.9cm}p{2.3cm}p{3.3cm}p{0.8cm}p{1.9cm}p{2.0cm}}
\noalign{\smallskip}
\multicolumn{7}{p{17.6cm}}{Table 1.  (continued) }\\
\hline \noalign{\smallskip}
Name(s)    & RA (J2000)       & Opt. Ctp. & V, $B-V$, $U-B$     & J & F$_{x}$  & P$_{orb}$(d)   \\
           & DEC (J2000)      &           & Sp. type, E$_{B-V}$ & H & $\mu$Jy  & P$_{pulse}$(s) \\
 Type      & $l^{II}, b^{II}$ & [FC]      &                     & K &          &                \\

\noalign{\smallskip}\hline \noalign{\smallskip}
IGR J16465$-$4507 & 16 46 35.26   &  &                & 10.537 & 16           &  \\
                  & $-$45 07 04.5 &  & B0.5I,         & 10.079 & (18--60 keV) & 228 \\
T, P              & 340.1, 0.13   &  & [560]     & 9.843  & [446]        & [447] \\
\\
\hline
\end{tabular}
\end{table*}

\clearpage
\begin{table*}[t]
\begin{tabular}{p{3.6cm}p{1.9cm}p{2.2cm}p{3.3cm}p{0.8cm}p{1.9cm}p{1.8cm}}
\noalign{\smallskip}
\multicolumn{7}{p{8.6cm}}{Table 1.  (continued) }\\
\hline \noalign{\smallskip}
Name(s)  & RA (J2000)       & Opt. Ctp. & V, $B-V$, $U-B$      & J & F$_{x}$  & P$_{orb}$(d)   \\
         & DEC (J2000)      &           & Sp. type, E$_{B-V}$  & H & $\mu$Jy  & P$_{pulse}$(s) \\
 Type    & $l^{II}, b^{II}$ & [FC]      &                      & K &          &               \\

\noalign{\smallskip} \hline \noalign{\smallskip}
XTE J1739$-$302   & 17 39 11.58    &             & 14.89, 2.97, 1.68  & 8.600 & 136         &       \\
IGR J17391$-$3021 & $-$30 20 37.6  &             & O8Iab(f), 3.4      & 7.823 & (2--25 keV) & 1500?   \\
T                 & 358.1, 0.45    &             & [562]              & 7.428 & [715]       & [684]  \\
\\
\hline
\end{tabular}
\end{table*}

\clearpage
\begin{table*}[t]
\begin{tabular}{p{3.3cm}p{1.9cm}p{2.3cm}p{3.3cm}p{0.8cm}p{1.9cm}p{2.0cm}}
\noalign{\smallskip}
\multicolumn{7}{p{17.6cm}}{Table 1.  (continued) }\\
\hline \noalign{\smallskip}
Name(s)      & RA (J2000)        & Opt. Ctp. & V, $B-V$, $U-B$     & J & F$_{x}$    & P$_{orb}$(d)   \\
             & DEC (J2000)       &           & Sp. type, E$_{B-V}$ & H & $\mu$Jy    & P$_{pulse}$(s) \\
 Type        & $l^{II}, b^{II}$  & [FC]      &                     & K &            &                \\

\noalign{\smallskip}  \hline \noalign{\smallskip}
4U 1807$-$10   & 18 11         &            &         &     & $<$2--10   &   \\
               & $-$10 54      &            &         &     &            &   \\
T              & 18.6, +3.9    &            &         &     & [215]      &   \\
\\
\hline
\end{tabular}
\end{table*}

\clearpage
\begin{table*}[t]
\begin{tabular}{p{3.3cm}p{1.9cm}p{2.3cm}p{3.3cm}p{0.8cm}p{1.9cm}p{2.0cm}}
\noalign{\smallskip}
\multicolumn{7}{p{8.6cm}}{Table 1.  (continued) }\\
\hline
\noalign{\smallskip}
Name(s)   & RA (J2000)        & Opt. Ctp.  & V, $B-V$, $U-B$      & J & F$_{x}$ & P$_{orb}$(d)   \\
          & DEC (J2000)       &            & Sp. type, E$_{B-V}$  & H & $\mu$Jy & P$_{pulse}$(s) \\
 Type     & $l^{II}, b^{II}$  & [FC]       &                      & K &         &                \\

\noalign{\smallskip} \hline \noalign{\smallskip}
GS 1839$-$06  & 18 41 42     &       &     &     & 1        &   \\
              & $-$05 51     &       &     &     &          &   \\
T             & 26.6, $-$0.5 &       &     &     & [399]    &   \\
\\
\hline
\end{tabular}
\end{table*}

\clearpage
\begin{table*}[t]
\begin{tabular}{p{3.6cm}p{1.9cm}p{2.1cm}p{3.3cm}p{0.8cm}p{1.9cm}p{1.90cm}}
\noalign{\smallskip}
\multicolumn{7}{p{8.6cm}}{Table 1.  (continued) }\\
\hline
\noalign{\smallskip}
Name(s)  & RA (J2000)       & Opt. Ctp. & V, $B-V$, $U-B$     & J & F$_{x}$ & P$_{orb}$(d)   \\
         & DEC (J2000)      &           & Sp. type, E$_{B-V}$ & H & $\mu$Jy & P$_{pulse}$(s) \\
 Type    & $l^{II}, b^{II}$ & [FC]      &                     & K &         &                \\

\noalign{\smallskip}\hline \noalign{\smallskip}
XTE J1859+083  & 18 59.1   &   &  &  & 8       &       \\
               & 08 15     &   &  &  &         & 9.801 \\
T, P           & 41.1, 2.1 &   &  &  & [474]   & [474] \\
\\
\hline
\end{tabular}
\end{table*}

\clearpage
\begin{table*}[t]
\begin{tabular}{p{3.3cm}p{1.9cm}p{2.3cm}p{3.3cm}p{0.8cm}p{1.9cm}p{2.0cm}}
\noalign{\smallskip}
\multicolumn{7}{p{17.6cm}}{Table 1.  (continued) }\\
\hline \noalign{\smallskip}
Name(s)    & RA (J2000)       & Opt. Ctp. & V, $B-V$, $U-B$     & J & F$_{x}$  & P$_{orb}$(d)   \\
           & DEC (J2000)      &           & Sp. type, E$_{B-V}$ & H & $\mu$Jy  & P$_{pulse}$(s) \\
 Type      & $l^{II}, b^{II}$ & [FC]      &                     & K &          &                \\

\noalign{\smallskip}\hline \noalign{\smallskip}
IGR J19140+0951   & 19 14 04.2    &       &         & 8.55 & 75--100     & 13.558 \\
IGR J19140+098    & 9 52 58.3     &       &         & 7.67 & (1--20 keV) &        \\
                  & 44.3, $-$0.46 & [344] & [344]   & 7.06 & [670]       & [140] \\
\\
\hline
\end{tabular}
\end{table*}

\clearpage
\begin{table*}[t]
\begin{tabular}{p{3.7cm}p{1.9cm}p{2.1cm}p{3.3cm}p{0.8cm}p{1.8cm}p{1.9cm}}
\noalign{\smallskip}
\multicolumn{7}{p{8.6cm}}{Table 1.  (continued) }\\
\hline \noalign{\smallskip}
Name(s)     & RA (J2000)       & Opt. Ctp. & V, $B-V$, $U-B$     & J & F$_{x}$ & P$_{orb}$(d)   \\
            & DEC (J2000)      &           & Sp. type, E$_{B-V}$ & H & $\mu$Jy & P$_{pulse}$(s) \\
 Type       & $l^{II}, b^{II}$ & [FC]      &                     & K &         &               \\

\noalign{\smallskip} \hline \noalign{\smallskip}
4U 2030+40  & 20 32 26.5       & V1521 Cyg      &               & 15.309 & 90--430     & 0.20     \\
Cyg X-3     & +40 57 09        &                & WR star, 6.3  & 13.192 &             &          \\
R           & 79.8, +0.7       & [812]          & [792]         & 11.921 & [64]        & [590]    \\
\\
\hline
\end{tabular}
\end{table*}

\end{document}